\documentclass[12pt]{article}
\usepackage{amsmath,amssymb,amsthm,amsfonts}
\usepackage{amsmath,amssymb,makeidx,amsfonts,latexsym,enumerate}

\title{Interference of probabilities and number field structure of quantum models}
\author{Andrei Khrennikov\\
International Center for Mathematical\\
Modeling in Physics and Cognitive Sciences,\\
MSI, University of V\"axj\"o, S-35195, Sweden\\
Email:Andrei.Khrennikov@msi.vxu.se}
\begin{document}
\maketitle

\begin{abstract}
We study the probabilistic consequences
of the choice of the basic number field in quantum formalism.
We demonstrate that by choosing a number field for a linear space representation
of quantum model it is possible to describe various interference phenomena.
We analyse interference of probabilistic alternatives induced by real, complex,
hyperbolic (Clifford) and $p$-adic representations. 
\end{abstract}

\section{Introduction}
In this paper we study the role of the basic number field, namely number field that can 
be used for the linear space representation, in quantum formalism. Since the creation of
quantum formalism, there are continuous debates on the role of the  number filed structure of
quantum theory, see, for example, [1]-[14]. There is the large diversity of opinions on this problem. 

We just point out that many consider the use of complex numbers as the convenient mathematical description
of two dimensional real quantities. From this point of view $z=x+iy\in {\bf C}$ is just the vector $v=(x,y)\in {\bf R}^2.$ 
This viewpoint justifies the reduction of complex Shr\"odinger's 
equation to the pair of real equations, in hydrodynamic models of quantum theory as well 
as the pilot wave theory [15], [16]. It is important to note that, in fact, the real hydrodynamic model was one of the first quantum models.
In principle, we can consider Shr\"odinger's wave mechanics as the complexification of the hydrodynamic model.
In this context it is natural to ask ourself: What was the main advantage of this complexification? 
I think it was the {\bf{linearization}} of nonlinear hydrodynamic equations.

Our investigation on the role of number field in quantum formalisms is strongly based of famous 
Dirac's analysis [1] of the foundations of quantum theory. Regarding to the appearance of 
complex numbers in quantum formalism, Dirac noted that by using only real coefficients for 
superpositions of states we could not describe all superpositions that can be produced 
in experiments with elementary particles. This remark is also very important for us.

In the present paper we find probabilistic roots of the appearance of the complex structure. 
It seems that the main aim of the introduction of complex numbers was the linearization (in a ${\bf C}$-linear 
space) of quantum rule for interference of probabilistic alternatives. We also consider the experimental 
domain that can be described by the ${\bf R}$-linearization. We found that the latter linearization describes only 
maximal interference (the $\cos\theta=\pm 1)$ and trivial interference
($\cos\theta=0).$

Then we study the `quantum formalism' for the algebra of so called hyperbolic numbers ${\bf G}$, 
namely two dimensional
Clifford algebra\footnote{We call this algebra hyperbolic algebra, because
hyperbolic functions are naturally related to the algebraic structure
of ${\bf G}$ through a hyperbolic generalization of Euler's formula for the complex
numbers.} 
(here $z=x+jy, x,y\in R$ and $j^2=1,$ see, e.g. [6], [14]).
Suddenly we discovered the possibility of a new type of interference, namely 
{\bf{hyperbolic interference.}} In the opposite to the standard quantum formalism in that 
the conventional probabilistic rule for the addition of alternatives is perturbed by $\cos \theta-$factor, 
in the hyperbolic case we get $\cosh \theta-$perturbation. It may be that such an `interference' could be 
observed in some experiments.

Finally, we provide probabilistic analysis of `quantum formalism' for the fields of $p$-adic numbers $ {\bf Q}_p$, 
see, e.g. [5], [7]-[14] 
for $p$-adic physical models. We recall that first
$p$-adic models appeared in string theory as an attempt to modify the description of space-time on 
Planck distances.\footnote{Of course, we did not forget about the work [5]. However, 
Beltrametti and Cassinelli obtained the following negative result: $p$-adic linear
space could not be the used for quantization in the framework of quantum logic. As a 
consequence, this paper did not stimulate further investigations in $p$-adic physics.}
There are two main types of $p$-adic quantum models:

(1) wave functions $\varphi: {\bf Q}_p\rightarrow {\bf C};$

(2) wave function $\varphi: {\bf Q}_p\rightarrow  {\bf Q}_p$ or into one of extensions of $ {\bf Q}_p.$ 

The first model is based on the representation in the standard ${\bf C}$-linear space, see [7]-[10]. 
There is nothing special (compare to the conventional quantum theory)
from the  probabilistic viewpoint. The second model is based on the representation 
in linear spaces over $ {\bf Q}_p$ or its extensions, see [11]. 
We consider probabilistic consequences of such representations.

We found that from the probabilistic viewpoint
the $p$-adic quantum formalism describes a rather special
sub-domain of standard quantum theory (over the field of complex numbers).
For some physical models it might be more fruitful to use $p$-adic quantization, because
the $p$-adic geometry might be more natural to describe such phase behaviour.
From this point of view we discuss a $p$-adic analogue of the interference experiment.
Such an experiment would exhibit unusual
fluctuations of the spatial distribution of probability: 
it should fluctuate discontinuously with respect to the ordinary
Euclidean geometry (but continuously with respect to the $p$-adic topology), see Example
on the $p$-adic interference model.

Of course, in the present note we do not plan to provide  extended analysis of the use of various
algebraic structures in physics (especially quantum). We would like just to mention investigations
on quaternionic models, e.g., [17], [18], [3], [6], $p$-adic models, e.g. [5], [7]--[14], 
models over supercommutative Banach (and more general
topological) superalgeras,e.g. [12]--[14], applications of Clifford algebra, e.g. [6].

\section{Probabilistic roots of complex and real linear representations}
It is well known that the creation of quantum formalism was, in fact, induced by two kinds of experiments
(see, for example, the introductory chapters of Dirac's  or Heisenberg's books,  [1] or [19]):

1) {\bf Corpuscular}: Black body radiation, photoelectric effect,...

2) {\bf Wave}: Diffraction, two slit experiment for elementary (especially massive) particles,...

The first class of experiments showed that light has the corpuscular structure. 
Corpuscles of light, photons, demonstrated discrete behaviour similar to behaviour of macro objects,
for examples, balls. In such a situation it was natural to assume that statistical properties of these 
corpuscular objects would be described by the conventional probability theory.
\footnote{We remark that, in fact, in 20th the mathematical foundations of probability theory 
(Kolmogorov's axiomatics, 1933, [20]) were not yet created. 
There was the large diversity of approaches to the notion of probability in the mathematical as well 
physical communities.}
In particular, we should get the conventional rule for the addition of probabilities of alternatives:
\begin{equation}
\label{R1}
P=P_1+P_2.
\end{equation}
However, the two slit experiment implies that conventional rule ({\ref{R1}}) is violated. The right hand 
side of quantum rule for addition of alternatives contains a perturbation term:
\begin{equation}
\label{R2}
P=P_1+P_2+2\sqrt{P_1P_2}\cos \theta.
\end{equation}
Rules ({\ref{R1}}) and ({\ref{R2}}) are used as prediction rules (especially in more
general situation related to the formula of 
total probability, e.g.,  [21], 
see further considerations): if we know probabilities $P_1$ and $P_2$, then we can 
find (predict) probability $P.$ 

We recall that trigonometric behaviour of `quantum probabilities' was first observed
in interference experiments. One of the aims of quantum formalism was to derive
probabilistic transformation (\ref{R2}). In all textbooks on quantum mechanics 
transformation (\ref{R2}) is obtained by using calculations in the complex Hilbert
space. Well, such an approach reproduce the true experimental result, namely (\ref{R2}).
However, this ${\bf C}$-Hilbert space derivation of the `quantum rule' of the addition
of probabilities does not explain the origin of the linear space structure. In particular,
in such a way we could never get the answer to the following fundamental question:

"Why do we use complex numbers as the basis of linear calculus?" 

Moreover, the derivation of (\ref{R2}) on the basis of the ${\bf C}$-linear calculus
and not on the basis of the conventional probability theory (Kolmogorov's axiomatics,1933,
[20]) induces a rather common opinion that there is the crucial difference between 
the conventional (`classical') and quantum probabilities. In any case the majority of 
quantum and probabilistic communities is sure that interference rule (\ref{R2}) could never be derived
in the conventional probabilistic framework.

So, we shall study simultaneously two closely connected problems:

(A) Is it possible to derive quantum rule (\ref{R2}) for addition of 
probabilities of alternatives in the conventional probabilistic framework?

(B) Is it possible to derive the ${\bf C}$-linear space structure starting with
quantum rule (\ref{R2})?

The problem (A) was investigated in author's works [22], [23]. It was demonstrated
that (\ref{R2}) could be obtained in the conventional probabilistic framework. We shall
briefly discuss this derivation in section 3. We now start to investigate the problem (B).
At the moment we are not interested in the `classical' theoretical derivation of (\ref{R2}).
Quantum rule (\ref{R2}) is considered as just an experimental fact. We would like to derive 
the ${\bf C}$-linear structure (${\bf C}$-linear space probabilistic calculus based
on the superposition principle) starting with (\ref{R2}). In particular, such a derivation 
will provide the clear probabilistic explanation of the choice of the field ${\bf C}$
as the basic number filed for quantization.

We note that, in the opposite to the standard rule, 
the quantum rule is given by the {\bf{nonlinear}} transformation:
\[f_{\rm{sq}}(P_1, P_2)=P_1+P_2+2\sqrt{P_1P_2}\cos\theta \;.\]
(for $\cos\theta \not=0.)$
We can ask: 

"Is it possible to linearize this transformation?" 

We first consider the case of maximal interference, namely $\lambda=\cos\theta=\pm 1.$
Here
\[
f_{\rm{sq}}(P_1, P_2)=P_1 \pm 2\sqrt{P_1P_2}+P_2=\vert \sqrt{P_1}+\epsilon\sqrt{P_2}\vert^2, \epsilon=\pm 1.
\]
Therefore, instead of the nonlinear transformation, we can work with the linear transformation 
for square roots of probabilities:

$g(x_1, x_2)=x_1 + \epsilon \; x_2,$ where $x_1=\sqrt{P_1}, x_2=\sqrt{P_2}.$

It seems that there is no special physical meaning 
in this linearization. It is just a question of better mathematical 
representation of the quantum probabilistic transformation (in the particular case, $\lambda=\pm 1).$

In the case $\lambda=\cos\theta\not = 0, \pm 1$ we cannot provide the linearization over ${\bf R}.$ 
However, such a linearization can be easily performed over ${\bf C}$:
\begin{equation}
\label{P}
f_{sq}(P_1, P_2)=|\sqrt{P_1} + \epsilon \;\sqrt{P_2}|^2,\;\mbox{where}\; \epsilon= e^{i\theta}\;.
\end{equation}
This explains why we use ${\bf C}$-linear formalism and not 
${\bf R}$-linear. This also implies that it seems that the only source of the appearance 
of complex numbers in quantum formalism was the linearization of the transformation $f_{\rm{sq}}(P_1, P_2).$

This is the good place to present the following (rather long) citation from
Dirac's book [1]: `... {\small from the superposition of two given translation states for a
photon a twofold infinity of translational states may be obtained, the general one of which
is described by two parameters, which may be taken to be the ratio of the amplitudes of the two wave functions
that are added together and their phase relationship. This confirmation shows the need for allowing 
complex coefficients in equation (1). \footnote{The equation for the superposition of two states: 
$c_1\vert A>+ c_2\vert B>=\vert R>.$}  
If these
coefficients were restricted to be real, then, since only their ratio is of importance for determining the 
direction of the resultant ket vector $\vert R>$ when $\vert A>$ and $\vert B>$ are given, there would be only
a simple infinity of states obtainable from the superposition.'}

Our considerations imply that we have to observe  (in some experiment) a new transformation
of probabilities different from quantum rule ({\ref{R2}}) (for the addition of
probabilistic alternatives)  to start to use new linear representations over fields, algebras, rings
that are different from the field of complex numbers.

However, we can also use another way for investigations. We can study different nonconventional 
linear representations (over fields, algebras, rings) 
and probabilistic transformations induced by such representations. Then 
we should look for experiments that could produce
such new probabilistic transformations.

Before to go to such nonconventional models, we demonstrate why the linearization of transformation
({\ref{R2}}) is so useful for quantum calculations. 

Typically in statistical experiments in classical physics we do not use just the 
trivial addition of alternatives, ({\ref{R1}}). We apply so called {\it formula of total probability} 
based on Bayes' formula for conditional probabilities, see, e.g., [21]. 
We present this formula in the situation that is natural for further quantum considerations.

Let $a=a_1, a_2$ and $b=b_1, b_2$ be two dichotomic ("classical") physical variables. Then we have:
\begin{equation}
\label{TP}
P(a=a_j)=P(b=b_1)P(a=a_j|b=b_1)+P(b=b_2)P(a=a_j|b=b_2),
\end{equation}
where conditional probabilities are given by 
Bayes' formula $(i,j=1,2):$

$P(a=a_j|b=b_i)=\frac{P((a=a_j)\wedge (b=b_i))}{P(b=b_i)}.$

In experiments with elementary particles, instead of the conventional 
formula of total probability, we have to use another formula\footnote{First this was observed as just
the experimental fact in interference experiments.}:
\begin{equation}
\label{R3}
p^a_j=p^b_1 p^{a/b}_{1j}+p_2^b p_{2j}^{a/b} + 2 \sqrt{p^b_1p_{1j}^{a/b}p_2^bp_{2j}^{a/b}}\cos\theta_j.
\end{equation}
Here $p_j^a$ and $p_j^b$ are probabilities to observe $a=a_j$ and $b=b_j,$ respectively,
for some state $\varphi; p_{ij}^{a/b}$ are probabilities to observe $a=a_j$ for the state determined by the condition $b=b_i.$ It must be noticed that formula ({\ref{R3}}) is just the experimental rule that had been found on the basis of the two slit experiment.
\footnote{We consider the discrete version of the two slit experiment. The variable $a$ 
describes discretized position of a particle on the screen that is used for the registration:
the screen is divided into two domains, $D_1$ and $D_2$ and $a=a_j$ if a particle is found in the domain
$D_j.$ The variable $b$ describes the following selections (filters):
$b=b_1:$ the first slit is open and the second one is closed;
$b=b_2:$ the inverse case. So, for example, the probability $p_{1j}^{a/b}$ is the probability 
to find $a=a_j$ if the first slit open and the second closed.
Probabilities $p_i^b, i=1,2,$ gives the probability for a particle to choose the 
$i$th slit (in the case in that both slits are open).
In fact, we have to be more careful: by the standard interpretation of quantum mechanics we could not determine
the concrete slit (when both slits are open) that is passed by
an individual particle. We can say that probabilities $p_i^b$ describe 
the distribution of slits with respect to the source of particles. In the symmetric case $p_1^b=p_2^b=1/2.$}

We now would like to linearize transformation ({\ref{R3}}). We do this as usual:

$p_j^a=|\sqrt{p_1^b}\sqrt{p_{1j}^{a/b}}+ e^{i\theta_j}\;\sqrt{p_2^b}\sqrt{p_{2j}^{a/b}}|^2.$

We can observe that this is a linear transformation with respect to square roots of probabilities: 

$ y_j=\sum_i x_i d_{ij},$

where $x_j=\sqrt{p_j^b}, y_j=\sqrt{p_j^a}$ and the matrice $D=(d_{ij}= e^{i\theta_j} \sqrt{p_{ij}^{a/b}}).$

Of course, this linear transformation for square roots of probabilities can be easily derived
by using linear space calculation (see any textbook on quantum mechanics);
by considering the following expansions
of state vectors:

$\varphi=e^{i\xi_1}\;\sqrt{p_1^{b}}|b_1>+ e^{i\xi_2}\;\sqrt{p_2^b}|b_2>;$

$\varphi= e^{i\eta_1}\;\sqrt{p_1^{a}}|a_1>+ e^{i\eta_2}\; \sqrt{p_2^a}|a_2>;$

$|b_1>=e^{i\xi_{11}}\;\sqrt{p_{11}^{a/b}}|a_1>+e^{i\xi_{12}}\;\sqrt{p_{12}^{a/b}}|a_2>$

$|b_2>=e^{i\xi_{21}}\;\sqrt{p_{21}^{a/b}}|a_1>+ e^{i\xi_{22}}\;\sqrt{p_{22}^{a/b}}|a_2>$

(here phases $\theta_j$ are easily computed via phases $\xi_{ij}$ and $\xi_j$).

It seems that quantum formalism, namely ${\bf C}$-linear space representation of 
probabilistic transformation (\ref{R3}), was created in such a way 
(probably this pathway was not recognized consciously). We underline again that 
({\ref{R3}}) was first found in interference experiments and only then the ${\bf C}$-linear 
representation of interference transformation (`quantization') was created to justify ({\ref{R3}}).

\section{Classical probabilistic derivation of the interference rule; trigonometric and
hyperbolic interference}
We start with the following contextual explanation of the violation of the classical
rule of the addition of probabilistic alternatives in quantum experiments. In fact, the classical
rule for addition of probabilities  (\ref{R1}) (as well as the formula of total probability (\ref{TP}))
can be only derived if all probabilities are defined for one fixed
complex of conditions (context), ${\cal C}$ (one fixed Kolmogorov probabilistic space, see, e.g., [21]).
So, we must have  $P=P(A\;/{\cal C})$ and $P_j=P(A_j\;/{\cal C}), j=1,2,$ where ${\cal C}$
is one fixed context.
However, in  quantum experiments corresponding to the superposition of states we 
use a few distinct contexts. Here $P=P(A\;/{\cal C})$ and $P_j=P(A_j\;/{\cal C}_j), j=1,2,$
where ${\cal C}$ and ${\cal C}_j$ are distinct complexes of physical conditions corresponding
to three different experimental frameworks.\footnote{For example, in the two slit experiment
we have: ${\cal C}$ - both slits are open, ${\cal C}_j$ only $j$th slit is open.}
Thus there is nothing surprising that in such a framework we get the violation of 
(\ref{R1}). This explanation of the violation of (\ref{R1}) could be found  already in the
book of  Heisenberg [19], see also Accardy [24], Ballentine [25], De Muynck [26]. 
Unfortunately,
these  works do not provide the classical probabilistic derivation of `quantum rule'
(\ref{R2}).\footnote{For example, Heisenberg explained in the details (and Accardy, Ballentine, De Muynck
and many other `contextual people' even in more details)
experimental dependence of probabilities obtained via the superposition principle. However, we ´could not find
even a trace of the classical probabilistic derivation. Moreover, it seems that Heisenberg
was sure that it would be impossible to derive (\ref{R2}) without wave particle dualism.}
Such a derivation was presented in author's papers [22], [23]. Here we consider
briefly the main points of this derivation.

We always can write:

$P=P_1+P_2+\delta,$ where $\delta =P-(P_1+P_2).$

We make the following normalization of the perturbation $\delta:$
\[\delta=2\sqrt{P_1P_2}\lambda,\]
where 

$\lambda=\delta /2\sqrt{P_1P_2}$ 

describes normalized statistical deviations induced by the transition from the original context 
${\cal C}$ to contexts ${\cal C}_j, j=1,2,$, see [22], [23] 
for the details. Thus, in general,
\begin{equation}
\label{R4}
P=P_1+P_2+2\sqrt{P_1P_2}\lambda
\end{equation}
We make the trivial (but fundamental) remark. There are only two possibilities:

1) $|\lambda|\leq 1$ and 2) $|\lambda|\geq 1$ 

(it is convenient to include $|\lambda|=1$ both in 1) and 2) to get the possibility
of the continuous transition from 1) to 2) and vice versa.

In the first case we can always represent
\begin{equation}
\label{R5}
\lambda=\cos \theta
\end{equation}
where $\theta$ is a kind of a phase parameter. So we get the trigonometric interference (\ref{R2}).
To derive the standard quantum formalism, we have to linearize probabilistic transformation
(\ref{R2}) in the complex Hilbert space, see the previous section.  

We underline one important consequence of our classical probabilistic derivation of the `quantum
probabilistic rule'. This rule should be obtained automatically in all experiments that produce
relatively small, $\vert \lambda\vert \leq 1,$ statistical deviations due to the transition 
from one context to another. In particular, if perturbations are negligibly small from the statistical
viewpoint, namely $\lambda=0,$ we get  `classical statistical physics' (one fixed context statistical physics).

After the above theoretical (classical probabilistic) derivation of quantum  transformation
(\ref{R2}), we could repeat considerations of section 2 and
${\bf C}$-linearize (\ref{R2}), the special case of the general transformation (\ref{R4}).

Trigonometric parameterization ({\ref{R5}}) of the normalized statistical deviation is fruitful for a large class of experiments.
It seems that it is a consequence of the use of the Euclidean metric for the description of such experiments. 
In experiments with phase-shifts, e.g., in neuron interferometry [27], the parameter $\theta$
is the real Euclidean angle.

In principle, it may be that some experimental framework induces another natural parameterization
\[\lambda=u(s)\]
where $|u(s)|\leq 1$ and $s$ (generalized phase)
belong to some domain $O.$  In general it need not be a domain 
in the Euclidean space. However, we can always use parameterization ({\ref{R5}}) with 
\begin{equation}
\label{RQ}
\theta(s)= arccos \;u(s)
\end{equation}
and 
${\bf C}$-linearize the probabilistic transformation:
\begin{equation}
\label{TR}
P=P_1+P_2+2\sqrt{P_1P_2}u(s)=|\sqrt{P_1}+e^{i\theta(s)}\;\sqrt{P_2}|^2
\end{equation}

{\bf{Remark.}} The previous consideration of the possibility of the trigonometric parameteresation 
of general transformation (\ref{TR}) implies that if the statistical deviation $\lambda$ is relatively 
small, $|\lambda|\leq 1,$ we can always use quantization in a ${\bf C}$-linear space. However, 
it may occur in some experiments that parameteresation $\theta(s),$ see (\ref{RQ}),  is not naturally 
related to this experiment. In that case it would be practically impossible to found the $\theta-$picture
that gives usual trigonometric oscillations (for example, $u(s)$ is a continuous parameterization
on some space of parameters $s\in O$ and $\theta(s)$ is a discontinues function on $O).$
In such a case it would be more natural to work in the $s$-picture. It would be fruitful to construct new quantization (a linear calculus over noncomplex space) to describe this phenomenon.

In case 2) we can always represent
\begin{equation}
\label{R6}
\lambda=\pm \cosh \theta
\end{equation}
Here we get hyperbolic probabilistic rule:
\[P=P_1+P_2 \pm 2\sqrt{P_1P_2}\cosh \theta\]
We observe that this rule can be linearized over the (commutative) algebra of so called hyperbolic numbers, see appendix 1:
\[P=|\sqrt{P_1}\pm e^{j\theta}\;\sqrt{P_2}|^2,\; \mbox{where}\; j^2=1.\]

This ${\bf G}$-linearization is based on the hyperbolic analogue of the complex
formula:

$a^2 + b^2 \pm 2 a b \cos \theta= \vert a \pm b e^{i\theta}\vert^2,$

that we have used to linearize (\ref{R2}). The corresponding hyperbolic formula is

$a^2 + b^2 \pm 2 a b \cosh \theta= \vert a \pm b e^{j\theta}\vert^2.$

In  case  2) the conventional
formula of total probability is also perturbed hyperbolically:

$p_j^a=p_1^b p_{1j}^{a/b} + p_2^b p_{2j}^{a/b} + 2 \epsilon_j \sqrt{p_1^bp_2^b  p_{1j}^{a/b} p_{2j}^{a/b}} \cosh \theta_j, \epsilon_j=\pm 1.$

In the same way as in the trigonometric case (by using hyperbolic numbers instead of
complex) this transformation can be realized as a ${\bf G}$-linear transformation
with respect to square roots of probabilities.

{\bf Remark.}(Geometric images of trigonometric and hyperbolic interferences)

(T): Let $P(\theta)=P_1+P_2 \pm 2\sqrt{P_1P_2}\cos \theta.$ Suppose that $P(\theta)$
gives the probability of a particle to be registered on the circle of radius $r=\theta.$
This is a kind of idealized two slit experiment (in the real experiment probabilities
$P_j$ also depend on $r=\theta).$ We get the following geometric image of such an
interference. There are circles of maximal, $r_k= 2\pi k,$ and minimal $r_k= \pi k,$ brightness. Brightness
decreases and increases rather slowly (trigonometrically) between these maximums and minimums.

(Ha): Let $P_+(\theta)=P_1+P_2 + 2\sqrt{P_1P_2}\cosh \theta$ with the same as in (T) assumption:
$r=\theta.$  Here brightness is minimal in the center and increases very quickly (exponentially)
for $\theta$ varying from $\theta=0$ to $\theta=\theta_{\rm{max}},$ where the value $\theta_{\rm{max}}$
of the hyperbolic phase parameter is determined by the condition:

$P_1+P_2 + 2\sqrt{P_1P_2}\cosh \theta_{\rm{max}}=1.$

So, $\theta_{\rm{max}}= \ln[q_+ +\sqrt{q_+^2-1}],$ where
$q_+=\frac{1-P_1-P_2}{2 \sqrt{P_1 P_2}}.$

(Hb): Let $P_-(\theta)=P_1+P_2 - 2\sqrt{P_1P_2}\cosh \theta.$ Here brightness is maximal
in the center and decreases very quickly (exponentially)
for $\theta$ varying from $\theta=0$ to $\theta=\theta_{\rm{min}},$ where the value $\theta_{\rm{min}}$
of the hyperbolic phase parameter is determined by the condition:

$P_1+ P_2 - 2 \sqrt{P_1P_2} \cosh \theta_{\rm{min}}=0.$

So, $\theta_{\rm{min}} = \ln[q_- + \sqrt{q_-^2 -1}],$ where
$q_-=\frac{P_1 + P_2}{2 \sqrt{P_1 P_2}}.$

(Ha)+(Hb): By considering $P(\theta)= P_+(\theta)$ on some intervals and
$P(\theta)= P_-(\theta)$ on others we get an interference
picture of circles of different brightness that is similar to the standard interference.
The main difference is that in the hyperbolic case brightness increases and decreases
very rapidly (exponentially).

\section{$p$-adic probabilistic transformations}
Elementary facts about the fields $ {\bf Q}_p$ 
of $p$-adic numbers (where $p$ is a prime number) 
can be found in appendix 2, see also [21]. Let $x \in  {\bf Q}_p$ and $|x|_p \leq 1.$
Then $x$ can be uniquely represented in the form:
\[x=p^l\;\epsilon, \;|\epsilon|_p=1, \; l=0,1,2,...\]
The $\epsilon$ is called a $p$-adic unit.
Here $|x|_p = p^{-l}.$ 
This representation is an analog of the representation of $x \in {\bf R}$ 
as $x=|x|\;\epsilon, \;\epsilon=\pm 1,$ or $x\in {\bf C}$ as $x=|x|\epsilon, \;|\epsilon |=1$
(so $\epsilon= e^{i\theta}).$ The calculus in the $ {\bf Q}_p$-linear space produces the following rule 
for the addition of probabilistic alternatives:

$P=|\alpha_1 + \epsilon \;\alpha_2|_p^2,\; P_1=|\alpha_1|_p^2, \; P_2=|\alpha_2|_p^2.$

A). Let $P_1>P_2.$ By using the strong triangle inequality (see appendix 2) we get:

$P\equiv P_1.$ 

So, we can represent $P=P_1+P_2+2\sqrt{P_1P_2}\lambda,$
where $\lambda=- \frac{1}{2}\sqrt{\frac{P_2}{P_1}}.$

Thus

\begin{equation}
\label{RR}
- \frac{1}{2}<\lambda \leq 0.
\end{equation}

B). Let $P_1<P_2.$ Here $\lambda= -\frac{1}{2}\sqrt{\frac{P_1}{P_2}},$ so $\lambda$ satisfies to ({\ref{RR}}).

C). Let $P_1=P_2.$ So $\alpha_i=p^l\;\epsilon_i, i=1,2, 
|\epsilon_i|_p=1.$ So $P_1=P_2=\;p^{-2l}.$ 
Here $P= c P_1,$ where $c= |\epsilon_1+ \epsilon \epsilon_2|_p^2.$ 
We remark that (as a consequence of the strong triangle inequality)
\[0\leq c\leq 1.\] As \[\lambda=\frac{c}{2}-1,\] we get\[-1\leq \lambda \leq -\frac{1}{2}.\]

Therefore:

{\it $p$-adic quantum mechanics induces the standard quantum probabilistic rule,
({\ref{R2}}), for the restricted range of angles,}
$\frac{\pi}{2}\leq \theta \leq \pi.$

So, by using $p$-adic linear representations we do not have something new from the probabilistic
viewpoint. However, by using $p$-adic numbers we can get the description of rather special
behaviour of phase. It may that such a phase behaviour might be observed in some experiments.

{\bf  Example.} ($p$-adic two slit experiment). In such an experiment nonsymmetrical 
location of slits with respect to the "source" would imply just
the reductions of probability $P\rightarrow P_1$ (for $P_1>P_2),$ and  
$P\rightarrow P_2$ (for $P_2>P_1).$  In the symmetric case, $P_1=P_2=A,$
we would observe the following discrete fluctuations 
depending on the parameter $\epsilon$ belonging to the unit sphere $S_1(0)=\{x\in  {\bf Q}_p:|x|_p=1\}.$
In practical consideration the sphere $S_1(0)$ is reduced to the set 
 $N_p=\{1,2,\ldots, p-1, p+1, \ldots\}$ of natural numbers that are not divisible by $p.$ 
 
 We consider the simplest case $\epsilon_1=\epsilon_2=1.$ So $c=|1+\epsilon|_p^2.$ 
 Let $A= 1/p^{2l}.$
 We see that $P=c A=A$ for 
 $\epsilon = 1, \ldots, p-2;$ then $P=\frac{A}{p^2}$ for $\epsilon=p-1;$ 
 then again $P=A$ for $p=p+1, \ldots, 2p-2$ and $P=\frac{A}{p^2}$ 
 for $\epsilon=2p-1$ (if $p \not = 2$).
 Such fluctuations will be observed until $\epsilon$ approaches $p^2-1.$ 
 For $\epsilon=p^2-1, $ we get $P=\frac{A}{p^4.}$ 
 Then again $P=A$ for $\epsilon=p^2+1, \ldots, p^2+p-2, $ and $P=\frac{A}{p^2}$ 
 for $p=p^2+p-1$ and so on. 
 
 Thus there exist some exceptional
 values of the parameter $\epsilon$ that reduces the probability.
 Roughly speaking we have the following $p$-adic interference picture:
 there are portions of concentric circles (e.g., of radii $r_k= 1+\epsilon, \epsilon=k=1,2,....)$ 
 having the same brightness, between
 these portions periodically appear circles having essentially lower brightness;
 brightness is reduced due to divisibility of $r_k$ by powers of $p.$
 The main difference from the ordinary trigonometric interference is that brightness
 varies discontinuously depending on radius. However, we notice that this is discontinuity
 with respect to the Euclidean metric. In $p$-adic metric brightness behave continuously.
 
 We would like to mention one distinguishing feature of the
 $p$-adic quantum mechanics, ${\bf Q}_p$-linear probabilistic calculus.
 There exist representations of  canonical commutation relations by bounded
 operators in the $p$-adic Hilbert space, see [29], [30]. Therefore
 it might more convenient to provide ${\bf Q}_p$-linear
 calculations, instead of standard ${\bf C}$-linear calculations.
 In this way we could escape to work with unbounded operators.
 Moreover, if the hypothesis on the $p$-adic structure of space time on Planck
 distances be confirmed by further investigations (in the framework of string theory,
 [7]-[10] or general gravity, see, e.g., [31]), then the $p$-adic space representation
 of the corresponding probabilistic calculus would be very natural.

\section{Appendix 1: hyperbolic algebra}

A hyperbolic algebra {\bf{G}} is a two dimensional real 
algebra with the basis $e_0=1$ and $e_1=j, $ where $j^2=1.$ 
\footnote{So, it is the two dimensional  Clifford algebra.}
Elements of {\bf{G}} have the form $z=x + j y, \; x, y \in {\bf{R}}.$ 
We have $z_1 + z_2=(x_1+x_2)+j(y_1+y_2)$ and $z_1 z_2=(x_1x_2+y_1y_2)+j(x_1y_2+x_2y_1).$ 
This algebra is commutative. We introduce the involution in {\bf{G}} by setting 
$\bar{z} = x - j y.$ 
We set $|z|^2=z\bar{z}=x^2-y^2.$ 
We remark that  $|z|=\sqrt{x^2-y^2}$ is not well defined for an arbitrary $z\in {{\bf{G}}}.$ 
We set ${{\bf{G}}}_+=
\{z\in{{\bf{G}}}:|z|^2\geq 0\}.$ We remark that ${{\bf{G}}}_+$ 
is the multiplicative semigroup: 
$z_1, z_2 \in {{\bf{G}}}^+ \rightarrow z=z_1 z_2 \in {{\bf{G}}}_+.$ 
It is a consequence of the equality 

$|z_1 z_2|^2=|z_1|^2 |z_2|^2.$

Thus, for $z_1, z_2 \in {{\bf{G}}}_+,$ 
we have $|z_1 z_2|=|z_1||z_2|.$ We introduce 
$$
e^{j\theta}=\cosh\theta+ j \sinh\theta, \; \theta \in {\bf{R}}.
$$ 
We remark that 
$$
e^{j\theta_1} e^{j\theta_2}=e^{j(\theta_1+\theta_2)}, \overline{e^{j\theta}} 
=e^{-j\theta}, |e^{j\theta}|^2= \cosh^2\theta - \sinh^2\theta=1.
$$

Hence, $z=\pm e^{j\theta}$ always belongs to ${{\bf{G}}}_+.$ 
We also have 

$\cosh\theta=\frac{e^{j\theta}+e^{-j\theta}}{2}, \;\;\sinh\theta=\frac{e^{j\theta}-e^{-j\theta}}{2 j}\;.$

We set ${{\bf{G}}}_+^*=
\{z\in{{\bf{G}}}_+:|z|^2>0 \}. $ 
Let  $z\in {{\bf{G}}}_+^*.$  We have 

$z=|z|(\frac{x}{|z|}+j \frac{y}{|z|})= \rm{sign}\; x\; |z|\;(\frac{x {\rm{sign}} x}{|z|} +j\;
\frac{y {\rm{sign}} x}{|z|}).$

As $\frac{x^2}{|z|^2}-\frac{y^2}{|z|^2}=1,$  we can represent $x$ sign $x= \cosh\theta$ 
and $y$ sign $x=\sinh\theta, $ where the phase $\theta$ is unequally defined. 
We can represent each $z\in {{\bf{G}}}_+^*$ as 

$z = \rm{sign}\; x\;  |z|\; e^{j\theta}\;.$ 

By using this representation we can easily prove that ${{\bf{G}}}_+^*$
is the multiplicative group. Here $\frac{1}{z}=\frac{{\rm{sign}} x}{|z|}e^{-j\theta}.$ 
The unit circle in ${{\bf{G}}}$ is defined as $S_1 = \{z\in{{\bf{G}}}:|z|^2=1\}
=\{ z= \pm e^{j \theta}, \theta \in (-\infty, +\infty)\}.$ It is a multiplicative
subgroup of ${\bf G}_+^*.$

\section{Appendix 2: $p$-adic numbers}

The field of real numbers ${\bf R}$ is constructed as the completion of the
field of rational
numbers ${\bf Q}$ with respect to the metric $\rho(x,y)$ $=$ $\vert x - y
\vert$, where $\vert
\cdot\vert$ is the usual valuation of distance given by the absolute value.
The fields of $p$-adic
numbers ${\bf Q}_p$  are constructed in a corresponding way, but  using
other  valuations.
For a prime number $p$, the $p$-adic valuation $\vert \cdot\vert_p $ is
defined in the following
way. First we define it for natural numbers. Every natural number $n$ can
be represented as the
product of prime  numbers, $n$ $=$ $2^{r_2}3^{r_3} \cdots p^{r_p} \cdots$,
and we define
$\vert n\vert_p$ $=$ $p^{-r_p}$, writing $\vert 0 \vert_p$ $=0$  and
$\vert -n\vert_p$ $=$
$\vert n \vert_p$.  We then extend the definition of the $p$-adic valuation
$\vert\cdot\vert_p$
to all rational numbers by setting $\vert n/m\vert_p$ $=$ $\vert
n\vert_p/\vert m\vert_p$
for $m$ $\not=$ $0$. The completion of ${\bf Q}$ with respect to the metric
$\rho_p (x,y)$ $=$
$\vert x- y\vert_p$ is the locally compact field of $p$-adic numbers ${\bf
Q}_p$.

The number fields ${\bf R}$ and ${\bf Q}_p$ are unique in a sense, since by
Ostrovsky's
theorem (see [28], [10], [11]) $\vert\cdot\vert$ and  $\vert\cdot\vert_p$ are the
only possible  valuations
on ${\bf Q}$, but have quite distinctive properties. The field of real
numbers ${\bf R}$ with its
usual valuation satisfies $\vert n\vert$ $=$ $n$ $\to$ $\infty$ for
valuations of natural numbers $n$
and is said to be {\it Archimedean.\/} By a well know theorem of number
theory, [28], [10], [11], the only
complete Archimedean fields are those of the real and the complex numbers.
In contrast, the fields
of $p$-adic numbers, which satisfy $\vert n\vert_p$ $\leq$ $1$ for all $n$
$\in$ ${\bf N}$, are
examples of {\it non-Archimedean\/} fields.

Unlike the absolute value distance $\vert \cdot \vert$, the $p$-adic
valuation satisfies the strong
triangle inequality
\begin{equation}\label{str}
|x+y|_p \leq  \max[|x|_p,|y|_p],  \quad x,y \in {\bf Q}_p.
\end{equation}
Consequently the $p$-adic metric satisfies the strong triangle inequality
\begin{equation}
\label{s1}
\rho_p(x,y)\leq \max[\rho_p(x,z),\rho_p(z,y)], \quad x,y,z \in {\bf Q}_p,
\end{equation}
which means that the metric $\rho_p$ is an {\it ultrametric\/}.

Write $U_r(a)$ $=$ $\{x\in {\bf Q}_p: |x -a|_p \leq r\}$ and $U_r^-(a)$ $=$
$\{x\in {\bf Q}_p:
|x -a|_p < r\}$ where $r$ $=$ $p^n$ and $n$ $=$ $0$, $\pm 1$, $\pm 2$,
$\ldots$.
These are the ``closed'' and ``open''  balls in ${\bf Q}_p$ while the sets
$S _r(a)$ $=$ $\{x \in K: |x -a|_p = r \}$ are the spheres in ${\bf Q}_p$
of such radii $r$.
These sets (balls and spheres) have a somewhat strange topological
structure from the
viewpoint of our usual Euclidean intuition: they are  both open and closed at
the same time, and as such are called {\it clopen} sets.  Another
interesting property of $p$-adic
balls is that  two balls have nonempty intersection if and only if one of
them is contained
in the other. Also, we note that any point of  a $p$-adic ball can be
chosen as its center,
so such a ball is thus not uniquely characterized by its center and radius.
Finally, any $p$-adic
ball $U_r(0)$ is an additive subgroup of ${\bf Q}_p$, while the ball
$U_1(0)$ is also
a ring, which is called the {\it ring of $p$-adic integers} and is denoted
by ${\bf Z}_p$.

Any $x$ $\in$ ${\bf Q}_p$ has a unique  canonical expansion (which
converges in the
$\vert \cdot\vert_p$--norm) of the form
\begin{equation}
\label{a0}
x= a_{-n}/p^n +\cdots\ a_{0}+\cdots+ a_k p^k+\cdots
\end{equation}
where the $a_j$ $\in$ $\{0,1,\ldots, p-1\}$ are the ``digits'' of the
$p$-adic expansion. 

 This paper was prepared during my visit to University of Havanna. 
 I would like to thank Professor Lillian Alvarez for her hospitality.
 
 I would like to thank S. Albeverio, L. Accardy, L. Ballentine, E. Beltrametti, G. Cassinelli,
 W. De Baere, W. De Myunck, D. Greenberger,
 C. Fuchs, L. Hardy, T. Hida, P. Lahti, D. Mermin, A. Peres, I. Pitowsky, 
 A. Shiryaev, J. Summhammer, L. Vaidman and I. Volovich for fruitful discussions on 
 probabilistic foundations of quantum mechanics.

{\bf{References}}

[1] P. A. M.  Dirac, {\it The Principles of Quantum Mechanics.}
(Claredon Press, Oxford, 1995).

[2] J. von Neumann, {\it Mathematical foundations
of quantum mechanics.} (Princeton Univ. Press, Princeton, N.J., 1955).

[3] D. Finkelstein, J.M. Jauch, S. Schiminovich and D. Speiser, Foundations of 
quaternion quantum mechanics. {\it J. Math. Phys.}, {\bf 3}, 207 (1962).

[4] E. Beltrametti  and G. Cassinelli, {\it The logic of Quantum mechanics.}
(Addison-Wesley, Reading, Mass., 1981).

[5] E. Beltrametti  and G. Cassinelli, Quantum mechanics and $p$-adic
numbers.{\it Found. Phys.}, {\bf 2}, 1--7 (1972).

[6] K. Guerlebeck, W. Sproessig, {\it Quaternionic and Clifford Calculus for Physicists and Engineers.}
(J. Wiley and Sons, 1998).

[7] I. V. Volovich,  $p$-adic string, {\em Class. Quant. Grav.\/}
{\bf 4}  83--87 (1987).

[8]  P.\,G.\,O. Freund  and  E.~Witten, Adelic string amplitudes,
{\em Phys. Lett.~B\/} {\bf 199}  191--195 (1987).

[9] Frampton P. H. and Okada Y., $p$-adic string
$N$-point function. {\it Phys. Rev. Lett. B }, {\bf 60}, 484--486 (1988).

[10] V. S.~Vladimirov, I.\,V.~Volovich, E.I.~Zelenov,
  {\em $p$-adic numbers in
mathematical physics\/} (World Scientific Publ., Singapore 1994).

[11] A.~Yu.~Khrennikov,  {\em $p$-adic valued distributions in mathematical 
physics.} (Kluwer Academic Publishers, Dordrecht 1994).

[12] V. S. Vladimirov   and I. V. Volovich,  Superanalysis, 1. 
Differential Calculus.    {\it Teor. and Matem. Fiz.}, {\bf 59}, No. 1, 3--27 (1984).

[13] V.S. Vladimirov   and I. V. Volovich ,  Superanalysis, 2. 
Integral Calculus. {\it Teor. and Matem. Fiz.}, {\bf 60},
No. 2, 169--198 (1984).

[14] A. Yu. Khrennikov, {\it Supernalysis.}  (Kluwer Academic Publishers, 
Dordreht/Boston/London, 1999).

[15] D. Bohm, {\it Quantum theory, Prentice-Hall.} 
(Englewood Cliffs, New-Jersey, 1951)

[16] D. Bohm  and B. Hiley, {\it The undivided universe:
an ontological interpretation of quantum mechanics.}
(Routledge and Kegan Paul, London, 1993)

[17] J. von Neumann and G. Birkhoff, {\it Annals of Mathematics,}
{\bf 37}, 823 (1936).

[18] D. Finkelstein, J. M. Jauch, S. Schiminovich and D. Speiser,
{\it J. Math. Phys.,} {\bf 3,} 207 (1962).

[19] W. Heisenberg, {\it Physical principles of quantum theory.}
(Chicago Univ. Press, 1930).

[20]  A. N. Kolmogoroff, {\it Grundbegriffe der Wahrscheinlichkeitsrechnung.}
Springer Verlag, Berlin (1933); reprinted:
{\it Foundations of the Probability Theory}. 
Chelsea Publ. Comp., New York (1956);

[21]  A. N. Shiryayev, {\it Probability.} Springer, New York-Berlin-He (1984).

[22] A. Yu. Khrennikov, {\it Ensemble fluctuations and the origin of quantum probabilistic
rule.} Rep. MSI, V\"axj\"o Univ., {\bf 90}, October (2000).

[23] A. Yu. Khrennikov, {\it Linear representations of probabilistic transformations induced
by context transitions.} Preprint quant-ph/0105059, 13 May 2001.

[24] L. Accardi, {\it Urne e Camaleoni: Dialogo sulla realta,
le leggi del caso e la teoria quantistica.} Il Saggiatore, Rome (1997)
(English translation to be published by Kluwer Academic Publ.).

[25] L. Ballentine, {\it Interpretations of probability and quantum theory.}
To be published in {\it Proceedings of the Conference "Foundations of Probability and Physics",}
V\"axj\"o, Sweden-2000.

[26] W. M. De Muynck, {\it Interpretations of quantum mechanics,
and interpretations of violations of Bell's inequality.} To be published
in {\it Proceedings of the Conference "Foundations of Probability and Physics",}
V\"axj\"o, Sweden-2000.

[27] U. Bose and H. Rauch (Editors), {\it Neuron Interferometry.}
(Clarendon Press, Oxford, 1979).

[28] W. Schikhov, {\it Ultrametric Calculus}, Cambridge Univ. Press,
Cambridge (1984)

[29] S. Albeverio,  A.Yu. Khrennikov, {\it  J. of Phys. A},  {\it 29}, 5515--5527 (1996).

[30] A.Yu. Khrennikov, {\it Non-Archimedean analysis: quantum
paradoxes, dynamical systems and biological models.}
(Kluwer Acad. Publishers, Dordreht/Boston/London, 1997).

[31] I. Ya. Aref'eva, B. Dragovich, P. H. Frampton, I. V. Volovich, {\it Int. J. of 
Modern Phys., A}, {\bf 6}, No 24, 4341--4358 (1991).

\end{document}